\documentstyle [12pt]{article}
\begin {document}
\begin {titlepage}
hep-th/9501139
\centerline { PATH INTEGRAL APPROACH TO BOSONIZATION OF D=2 FIELD THEORIES}
\vspace{10mm}
\centerline{S.KIYANOV-CHARSKY}
\vspace{10mm}
\centerline{Department of Physics,Forestry Academy,}
\centerline{Institutsky per.5,St-Petersburg 194018,}
\centerline{Russia}
\vspace{10mm}

{\bf Abstract }                           \\
\vspace{5mm}
   We suggest a method of bosonizing any D=2 theory. We demonstrate
   how it works with the examples of the Thirring and the Schwinger
   models, known results are reproduced. This method, being applied to
   the Gross-Neveu model, yields nonlinear boson WZW-type theory with
   additional constraint in the field space. Relation to the nonlinear
   $\sigma$ - model is also discussed.

\vspace{20mm}
\end {titlepage}

{\bf 1 Introduction}
\vspace{5mm}

   Apart from supersymmetry approach, the equivalence between
bose and fermi forms of several D=2 theories is interesting for
many reasons. Over the last two decades, when constructing quantized
models of corresponding classical fermi theories, the
first step was to trasform these theories into bose form \cite{A}.
This is the case with the massive Thirring model, which is equivalent
to the sine-Gordon \cite{B}, the chiral Schwinger model with
its bose form \cite{C}, D=2 QCD \cite{D} and others \cite{E}.

  In what sense bose and fermi theories are equivalent? A standard
answer \cite{B,F} is that both these theories reproduce the same quantum
currents algebra or, in other words, if both theories give
identical correlation functions of bilinear fermion composite
operators. This paper proposes another approach: the bose form of
a given fermi model is another representation of it and, in fact,
is included in this fermi theory. So, we introduce a gauge group
valued field $\Omega$, corresponding to anomalous chiral transformation.
The anomalous subgroups of the full chiral gauge group $G$
play an exclusive role in our approach because they, and only they
can induce the effective boson action. By this method, we are able
to obtain the expressions for the nonabelian chiral densities in
terms of the boson field $\Omega$(see sec.4 below). This procedure
generalizes the similar one for abelian theories(see sec.3 and
ref. \cite{I}).

  Somewhat analogous approach to bosonization via path integrals
was developed by many authors \cite{G}. In these works, by introducing
the auxiliary boson fields, the expression for generating functional
was obtained in terms of the bose fields after the integrating
 over fermions. Later, this method became apply to the fermi
theories in higher dimensions, for example, to D=4 QCD or SQCD \cite{H}.

  In the recent years, a more straightforward method of D=2 bosonization
was developed for abelian theories \cite{I}, where the auxiliary boson
field was introduced through the chiral transformation.
But in these works, the full recipe of introducing such boson
field is absent. Our paper solves the problem of bosonization in a
general framework by including a gauge group valued auxiliary
field $\Omega$ to describe the bose degrees of freedom.

  The paper is organized as follows. The method is formulated in
sec.2, where we derive the appropriate bosonic action in a general
form. In sec.3 we apply this method to several known abelian cases -
the massive Thirring and the chiral Schwinger models, after that we
proceed (sec. 4) to the nonabelian case, represented by the U(N)
Gross-Neveu model. We shall find that this model is equivalent to
the nonlinear chiral WZW - like model with an additional field constraint.
This is a new result.Some remarks on the full bosonization
problem are scoped in conclusion.

  Throughout the paper we use the following notations: $\gamma^0=\sigma^1,
\gamma^1=i\sigma^2,\gamma^5=\sigma^3;g^{\mu\nu}=(+,-)$ so that
$\gamma_5\gamma_{\mu}=\epsilon_{\mu\nu}\gamma^{\nu}$,
$\epsilon^{01}=-\epsilon_{01}=1$.
We shall deal with a formulation of a given quantum theory in terms
of path integrals. Now we proceed to our bosonization method.           \\
\\
\vspace{10mm}

{\bf 2 A general method }
\vspace{5mm}

 Consider a renormalizable fermion model. This theory is quadratic
on fermion fields, or it can be achieved by introducing a set of
auxiliary fields. The action is
$$S_0=\int d^2x(\bar{\psi}\widehat{D}\psi)+S_{boson}.$$
The $\psi$ field belongs to the space of the fundamental representation
of the chiral gauge group $U_{L}(N)*U_{R}(N)$. The Dirac operator
$\widehat{D}$ depends on auxiliary fields and gauge fields $A_{\mu}$.
The generating functional for the Green's functions has the form
$$Z(J)=\int D\bar{\psi}D\psi Db e^{iS(J)}, \eqno(1)$$
where $S(J)=S_0+S_{sources}=
\int d^2x(\bar{\psi}\widehat{D}(J)\psi)+S_{boson}, b$
denotes all boson fields in a theory. Here, we have supposed that the
sources $J$ are introduced for fermion bilinears such as $\bar{\psi}\psi,
\bar{\psi}\gamma_{\mu}T^a\psi$,
$\bar{\psi}\gamma_{\mu}\gamma_5T^a\psi$ and others. That is because we are
interested in the correlation functions for currents in a theory, for
instance, the quantum currents algebra.

  Consider now a new functional $Z^1(J)$, which has the same Green's
functions for currents operators in terms of $\bar{\psi}$, $\psi$ fields
which $Z(J)$ has:
$$Z^1(J)=\int D\Omega Z(J), \eqno(2) $$
$D\Omega$ is a measure on the gauge group $G$, and  $\Omega$ is a function
of $x:$ $\Omega=\Omega(x) \in G$. We have introduced a new field $\Omega$
according to a given representation of a fermion wave function $\psi$,
$\Omega$ is an element of the chiral gauge group $G=U_L(N)*U_R(N)$.So the
quantum numbers of $\Omega$ are just the same as of the corresponding
gauge group parameters.

  After substitutions
$$\psi \to \Omega\psi,  \psi^+ \to \psi^+\Omega^+,$$
we obtain:
$$D\bar{\psi}D\psi \to e^{i\alpha_1(A,\Omega)}D\bar{\psi}D\psi,$$
$$\widehat{D}(J) \to \widehat{D}(J,\Omega)=
\tilde{\Omega}\widehat{D}(J)\Omega, \eqno(3)$$
$$\tilde{\Omega}=\gamma_0\Omega^+\gamma_0,$$
here $\alpha_1(A,\Omega)$ is a 1-cocycle on the gauge group \cite{J} (if
theory has no gauge fields, it is reduced to $\alpha_1(0,\Omega)$). In
fact $\alpha_1$ is the WZW action, including a topological part for the
nonabelian groups \cite{F,K}. From the general property of the 1-cocycle
$$\alpha_1(A^{\Omega},g)-\alpha_1(A,\Omega g)+\alpha_1(A,\Omega)=0,$$
we can deduce useful identities $\alpha_1(A^{\Omega}$, $\Omega^{-1})=
-\alpha_1(A,\Omega)$; $\alpha_1(A^{\Omega},1)=0.$ Now (2) has a form
$$Z^1(J)=\int D\Omega D b D\bar{\psi}D\psi
\exp(iS(J,\Omega)+iS_{boson}+i\alpha_1(A,\Omega)),$$
where
$$S(J,\Omega)=\int d^2x(\bar{\psi}\widehat{D}(J,\Omega)\psi),$$
and $\widehat{D}(J,\Omega)$ is given by (3).

  After integrating over the fermi fields we finally have a bosonic form
of the theory (1):
$$Z^1(J)=\int D\Omega D b \exp(\frac{i}{2}
{\rm Tr}\log\widehat{D}\widehat{D}^+(J,\Omega)+
iS_{boson}+i\alpha_1(A,\Omega)). \eqno(4)$$

  So for any given fermion theory (1) we are able to construct a
bose theory (4), both these theories have the same correlation
functions for currents and hence, are equivalent quantum mechanically.
This is the main result of this paper. Take into account that it is
necessary to do the Wick rotation to define
${\rm Tr}\log\widehat{D}\widehat{D}^+$ in (4) on euclidian compactified
space $T^2$.

  Note that instead of (2), one can consider a more general functional
$Z^1_{\rho}(J)=\int D\Omega \rho(\Omega)Z(J)$, where $\rho(\Omega)$ is an
arbitrary weight functional $\rho: G \to {\bf R}$, but we don't discuss
here an ambiguity of this kind.

  Now, turn to some D=2 fermion models and their bosonized versions. \\
\\
\vspace{10mm}
{\bf    3  Two abelian models}
\vspace{5mm}

  Starting from an abelian case, consider the Thirring model \cite{B}
$$S(J)=\int d^2x(\bar{\psi}(i\widehat{\partial}-m)\psi-
\frac{1}{2}g\bar{\psi}\gamma_{\mu}\psi\bar{\psi}\gamma^{\mu}\psi+
J_{\mu}\bar{\psi}\gamma^{\mu}\psi)=$$
$$\int d^2x(\bar{\psi}\widehat{D}(J)\psi)+S_{boson}, \eqno(5)$$
$$\widehat{D}(J)=i\widehat{\partial}-g\widehat{a}+\widehat{J}-m , $$
$$S_{boson}=\frac{g}{2}\int d^2x a_{\mu}a^{\mu}.$$
In this case we can't choose the group element $\Omega=e^{i\alpha(x)}$
because $\alpha(x)$ here is a phase of the wave function $\psi(x)$, and
it is unobservable. Instead of it, we use a fact, that in the massless
limit $m \to 0$, the Thirring model has a symmetry $U_L(1)*U_R(1)$. So,
we are able to choose $\Omega$ as an element of the coset
$U_L(1)*U_R(1)/U(1)$, $\Omega=e^{i\gamma_5\alpha(x)}$ and
$$\widehat{D}(J,\Omega)=
i\widehat{\partial}-\gamma_5(i\widehat{\partial}\alpha)-
g\widehat{a}+\widehat{J}-me^{2i\gamma_5\alpha},$$
due to (3). Then, we are going to calculate
${\rm Tr}\log\widehat{D}\widehat{D}^+(J,\Omega)$ using,
for example, the proper - time regularization method for elliptic
operator $\widehat{D}\widehat{D}^+(J,\Omega)$:
$${\rm Tr}\log\widehat{D}\widehat{D}^+(J,\Omega)=
-{\rm tr}\int^{\infty}_{\epsilon}\frac{dt}{t}
{\rm Tr} e^{-t\widehat{D}\widehat{D}^+(J,\Omega)}=$$
$$-\int^{\infty}_{\epsilon}\frac{dt}{t}
{\rm tr}\int d^2x\frac{d^2k}{(2\pi)^2}
e^{ikx}e^{-t\widehat{D}\widehat{D}^+(J,\Omega)}e^{-ikx}=$$
$$-\int^{\infty}_{\epsilon}\frac{dt}{t}
{\rm tr}\int d^2x\frac{d^2k}{(2\pi)^2}
\exp(-te^{ikx}\widehat{D}\widehat{D}^+(J,\Omega)e^{-ikx})=$$
$$-\int^{\infty}_{\epsilon}\frac{dt}{t}
{\rm tr}\int d^2x\frac{d^2k}{(2\pi)^2}
e^{-tk^2}\sum^{\infty}_{n=0}\frac{(-t)^n}{n!}(B_{\mu}k^{\mu}+C)^n=$$
$$\frac{A}{16\pi}{\rm tr}\int d^2x(4C-B_{\mu}B^{\mu})-
\frac{1}{16\pi}{\rm tr}\int d^2xB^{\mu}B^{\mu}+O(\epsilon), \eqno(6)$$
where $A \to \infty$ as $\epsilon \to 0$, ${\rm tr}$-trace over
$\gamma$-matrices;
$$C=\widehat{D}\widehat{D}^+(J,\Omega) ,
B_{\mu}=\gamma_{\mu}\widehat{D}^++\widehat{D}\gamma_{\mu},$$
so that ${\rm tr}(4C-B_{\mu}B^{\mu})=
-2{\rm tr}\gamma_{\mu}\widehat{D}\gamma^{\mu}\widehat{D}^+=
-4m^2{\rm tr} e^{4i\gamma_5\alpha(x)}=-8m^2\cos4\alpha(x),$
because in  D=2 for any vector
$P_{\mu}:\gamma_{\mu}\widehat{P}\gamma^{\mu}=0$.
Also in (6) we have used an operator identity
$e^A e^B e^{-A}=\exp(e^A B e^{-A})$.
The only term remained in (6) is
$${\rm tr}B_{\mu}B^{\mu}=
{\rm tr}(2\gamma_{\mu}\widehat{D}\gamma^{\mu}\widehat{D}^+)=$$
$$8(\partial_{\mu}\alpha\partial^{\mu}\alpha+
2g\epsilon^{\mu\nu}(\partial_{\nu}\alpha)a_{\mu}+g^2a_{\mu}a^{\mu}+
2\epsilon^{\mu\nu}(\partial_{\nu})J_{\mu}+J_{\mu}J^{\mu}+m^2\cos4\alpha).$$
After substitution of (6) in (4) and integrating over $a_{\mu}$ fields in
(4), one exactly gets sine-Gordon theory ( after rescaling
$\alpha \to \sqrt{\pi}\alpha$). Note, that the source $J_{\mu}$ in
(5) is coupled to the vector current in a fermionic form
$\bar{\psi}\gamma^{\mu}\psi$, while in boson theory it is coupled to the
same current in terms of bosonic field $\alpha(x)$, so we immediatly have
$$\bar{\psi}\gamma^{\mu}\psi=
\frac{1}{\sqrt{\pi}}\epsilon^{\mu\nu}\partial_{\nu}\alpha$$
as expected \cite{B,F}. Corresponding action turns out to be
$$S(J)=\int d^2x(\frac{Z_1}{2}\partial_{\mu}\alpha\partial^{\mu}\alpha+
\frac{1}{\sqrt{\pi}}\epsilon^{\mu\nu}(\partial_{\nu}\alpha)J_{\mu}+
m^2Z_m\cos4\alpha),$$
$$Z_1=(1+\frac{g}{\pi})^{-1} , Z_m=\frac{A+1}{2\pi}.$$

 The method of bosonization nonabelian Thirring model at 1/N expansion was
also considered in literature \cite{L}.
 The second interesting example is the chiral Schwinger model, widely
discussed by many authors \cite{A,C}, subject to quantization of an
anomaly theory. This model is described by the action
$$S_0=\int d^2x(\bar{\psi}\widehat{D}\psi-
\frac{1}{4}F_{\mu\nu}F^{\mu\nu}), \eqno(7)$$
$$\widehat{D}=i\widehat{\partial}+e\sqrt{\pi}\widehat{A}(1-\gamma^5)=
i\widehat{\partial}+2e\sqrt{\pi}\widehat{A}P_-.$$
The fermions belong to the representation space of the abelian group
$G=U_L(1)*U_R(1)$.Choosing
$\Omega=\exp(i\sqrt{\pi}(P_-\alpha+P_+\beta)) \in G$;
$P_{\pm}=\frac{1\pm\gamma_5}{2}$ -
the chiral projection operators such as $P_{\pm}\gamma_{\mu}=
\gamma_{\mu}P_{\pm}$, we obtain
$$\widehat{D}(\Omega)=\tilde{\Omega}\widehat{D}\Omega=
i\widehat{\partial}+\widehat{a}P_-+\widehat{b}P_+ ,$$
where
$$\widehat{a}=\sqrt{\pi}(2e\widehat{A}-\widehat{\partial}\alpha) ,
\widehat{b}=-\sqrt{\pi}\widehat{\partial}\beta ,$$
and in a way, similar to the eq.(6), we get
$${\rm Tr}\log\widehat{D}\widehat{D}^+(\Omega)=
-\frac{1}{16\pi}{\rm tr}\int d^2xB_{\mu}B^{\mu}+O(\epsilon),$$
with vanishing term ${\rm tr}(4C-B_{\mu}B^{\mu})$, and
$${\rm tr}B_{\mu}B^{\mu}=4{\rm tr}\widehat{D}\widehat{D}^+(\Omega)=
16\pi(\partial_{\mu}\alpha\partial^{\mu}\alpha+2eA_{\mu}(g^{\mu\nu}+
\epsilon^{\mu\nu})\partial_{\nu}\alpha+2e^2A_{\mu}A^{\mu}),\eqno(8)$$
where we have taken $\alpha=-\beta$. In deriving the eq.(8), it is
necessary to continue $\gamma_5$ matrix properly in ${\bf R^2}$,
because it changes its own hermitian property $\gamma^{+}_{5E}=
-\gamma_{5E}$, and $\gamma^2_{5E}=-1$, so one has an unusual
algebra of the projection operators: $P^{E}_{\pm}P^{E}_{\pm}=
\pm\gamma_{5E}, P^{E}_{\pm}P^{E}_{\mp}=1$ instead
of $P_{\pm}P_{\pm}=P_{\pm},P_{\pm}P_{\mp}=0$. After substitution of (8)
into (4) , one gets the bosonized chiral Schwinger model with a WZW term
\cite{C}.
\\
\vspace{10mm}

{\bf  4  Nonabelian model}
\vspace{5mm}

  Obviously, the nonabelian models are the most interesting since
they lead to nonlinear boson theories which can be quantized by the
Faddeev-Jackiw method \cite{M} for constraint theories. In this section we
are dealing with the U(N) Gross-Neveu model, where fermions form a
space of the fundamental representation of the group $G=U(N)$. We
shall see that this model leads to somewhat unusual boson theory.

  To begin with, let us consider a free-fermion model with the sources
$J_{\mu L},J_{\mu R}$ for the left and the right currents respectively,
coupled as follows: $2\bar{\psi}\widehat{J}_LP_-\psi$ and
$2\bar{\psi}\widehat{J}_RP_+\psi$;
$\widehat{J}_{L,R}=J^{a}_{\mu L,R}T^a\gamma^{\mu}$. Then,
$$S(J)=\int d^2x\bar{\psi}_i\widehat{D}^{ij}(J)\psi_j=$$
$$\int d^2x\bar{\psi}\widehat{D}(J)\psi,\eqno(9)$$
$$\widehat{D}(J)=i\widehat{\partial}+2\widehat{J}_LP_-+2\widehat{J}_RP_+,$$
and the element of the chiral symmetry group G is choosen to be
$\Omega^1=P_-\Omega+P_+\Omega^+ \in G=U_L(N)*U_R(N)$. So, for the matrix
operator $\widehat{D}(J,\Omega)=\tilde{\Omega^1}\widehat{D}(J)\Omega^1$
one finds
$${\rm tr}B_{\mu}B^{\mu}=16{\rm tr}_u((\partial_{\mu}\Omega)
(\partial^{\mu}\Omega^+)+
\frac{1}{2}\epsilon^{\mu\nu}(\partial_{\mu}\Omega)
(\partial_{\nu}\Omega)\Omega^{+^2}-$$
$$\frac{1}{2}\epsilon^{\mu\nu}(\partial_{\mu}\Omega^+)
(\partial_{\nu}\Omega^+)\Omega^2-\frac{i}{2}\epsilon^{\mu\nu}
J_{\mu L}\Omega(\partial_{\nu}\Omega^+)+\frac{i}{2}\epsilon^{\mu\nu}
J_{\mu R}(\partial_{\nu}\Omega)\Omega^+),\eqno(10)$$
where ${\rm tr}_u$-a trace over the U(N) group, while quadratic in sources
terms are omitted since they have not contributed to the correlation
functions in (4). The substitution of (10) into (4) gives the nonlinear
sigma model with a WZW term and a topological term related to
$\pi_2(S^2)$. Indeed, choosing a form of the element of U(2), for example,
is $\Omega=n_0I+i{\bf \sigma n}$, unitarity condition
$\Omega\Omega^+=1$ transforms into $n^2_i=1$, i=0,1,2,3, and
$${\rm tr}_u(\partial_{\mu}\Omega)(\partial^{\mu}\Omega^+)=
2\partial_{\mu}n_i\partial^{\mu}n_i,$$
$${\rm tr}_u\epsilon^{\mu\nu}(\partial_{\mu}\Omega)
(\partial_{\nu}\Omega)\Omega^{+^2}=4n_0\epsilon^{\mu\nu}
\epsilon_{abc}\partial_{\mu}n_a\partial_{\nu}n_bn_c .$$
So one gets the O(3) nonlinear sigma model, if $n_0=const$ \cite{N}.

  Corresponding action turns out to be
$$S_{free}=\frac{1}{\pi}\int d^2x
(\frac{1}{2}\partial_{\mu}n_a\partial^{\mu}n_a+
const\epsilon^{\mu\nu}\epsilon_{abc}\partial_{\mu}n_a\partial_{\nu}n_bn_c)+
S_{WZW} .\eqno(11)$$
The second term is a degree of the map $n_a(x):S^2 \to S^2.$
Note that from (10) and (4) one has the expressions for the currents
in bosonized theory such as
$$\bar{\psi}\gamma^{\mu}(1+\gamma_5)T^a\psi=
\frac{i}{2}\epsilon^{\mu\nu}{\rm tr}_u(T^a\partial_{\nu}\Omega\Omega^+) ,$$
$$\bar{\psi}\gamma^{\mu}(1-\gamma_5)T^a\psi=
-\frac{i}{2}\epsilon^{\mu\nu}{\rm tr}_u(T^a\Omega\partial_{\nu}\Omega^+).$$

  Now add the Gross-Neveu interaction term to a free theory (9):
$$S_{int}=\frac{g^2}{2}\int d^2x(\sum^{N}_{i=1}\bar{\psi}_i\psi_i)^2=
g\int d^2x\phi\bar{\psi}\psi-\frac{1}{2}\int d^2x\phi^2 .$$
It contributes to (11)
$$4g^2\phi^2{\rm tr}_u(\Omega^4+\Omega^{+^4}) ,\eqno(12)$$
so, for the boson theory one finds
$$S(\Omega)=S_{free}+
\frac{g^2}{4\pi}\int d^2x\phi^2{\rm tr}_u(\Omega^4+\Omega^{+^4})-
\frac{1}{2}\int d^2x\phi^2 .$$
After the integrating over $\phi$ field one obtains the nonlinear sigma
model action (11) with additional to $\Omega\Omega^+=1$ constraint from
(12): ${\rm tr}_u(\Omega+\Omega^+)^4-2{\rm tr}_u(\Omega+\Omega^+)^2=const$
as for $\Omega\in U(N)$ one has $\Omega^4+\Omega^{+^4}=
(\Omega+\Omega^+)^4-2(\Omega+\Omega^+)^2-2.$ This reduces the phase space
of the theory (11), giving a new result.
\\
\vspace{20mm}

{\bf 5 Conclusion.}
\vspace{5mm}

  In this paper, for any given fermion theory in D=2,
we have constructed the equivalent quantum mechanically bose theory,
which is described by the action (4). Physically it means, that theory
(4)is a nothing but the effective one for the fields, extracted from
the phase space of the fermion wave functions in some representation
of the internal symmetry group G. By this method, we have obtained the
bose theory for the Gross-Neveu model, which appeares to be a nonlinear
WZW - type model with a topological term and additional constraint
in the phase space. Note, that the appearence of the spin structure in
such a model was confirmed by the independent method \cite{O}.

  Many other applications of this explicit bosonization procedure can
be considered. For example, if one enlarges the internal symmetry group
to the conformal group, it can be possible to obtain the D=2 induced
gravity. It can be used also in D=4 theories for deducing the low-
energy effective actions.
\\
\vspace{10mm}

{\bf Acknowledgement}
\vspace{5mm}

  The author would like to thank Dr. Vassilevich D.V. for valuable
discussions, and for giving him preprint \cite{P}, where the problem of
non-abelian bosonization is solved by the different method.
\\

\end {document}